\documentclass[12pt,onecolumn]{emulateapj}

\usepackage{natbib}
\usepackage{graphicx}
\usepackage{psfig,graphics}
\usepackage{bbm}
\usepackage{epstopdf}
\newcommand{\beq}{\begin{equation}}
\newcommand{\eeq}{\end{equation}}

\newcommand{\gsi}{\,\raisebox{-0.13cm}{$\stackrel{\textstyle>}
{\textstyle\sim}$}\,}
\newcommand{\be}{\begin{equation}}
\newcommand{\ee}{\end{equation}}

\begin{document}

\title{Galaxies Correlating with Ultra-high Energy Cosmic Rays}

\vspace{0.25cm}
\author{Ingyin Zaw and Glennys R. Farrar}
\affil{Center for Cosmology and Particle Physics \&
Department of Physics\\ New York University, NY, NY 10003,USA}
\vspace*{-1.25cm}
\author{Jenny E. Greene\footnote{Hubble Fellow and Princeton-Carnegie Fellow}}
\affil{Department of Astronomy and Astrophysics
Peyton Hall\\ Princeton, NJ 08854,USA\\
\altaffiltext{2}{Hubble, Princeton-Carnegie Fellow}
}


\begin{abstract}
The Pierre Auger Observatory reports that 20 of the 27 highest energy cosmic rays have arrival directions within 3.2$^\circ$ of a nearby galaxy in the Veron-Cetty \& Veron Catalog of Quasars and Active Galactic Nuclei (12$^{\rm th}$ Ed.), with $\sim 5$ of the correlations expected by chance.  In this paper we examine the correlated galaxies to gain insight into the possible UHECR sources.  We find that 14 of the 21 correlated VCV galaxies are AGNs and we determine their bolometric luminosities.  The remaining 7 are primarily star-forming galaxies.  The bolometric luminosities of the correlated AGNs are all greater than $ 5 \times 10^{42} \,{\rm erg \,s^{-1}}$, which may explain the absence of UHECRs from the Virgo region in spite of the large number of VCV galaxies in Virgo, since most of the VCV galaxies in the Virgo region are low luminosity AGNs.  Interestingly, the bolometric luminosities of most of the AGNs are significantly lower than required to satisfy the minimum condition for UHECR acceleration in a continuous jet.  If a UHECR-AGN correlation is substantiated with further statistics, our results lend support to the recently proposed ``giant AGN flare" mechanism for UHECR acceleration.  
\end{abstract}

\section{Introduction}

The origin of ultra-high energy cosmic rays (UHECRs) has been an important question in astrophysics, for many decades.  Early observations \citep{AGASA1998} suggested a violation of the Grizsen-Zatsepin-Kuzmin (GZK) prediction that the UHECR energy spectrum must drop off at energies above $\sim 60 $ EeV. If confirmed, this would have been a major challenge for theory, but recent high-statistics observations by the HiRes and Pierre Auger collaborations find a downturn in the spectrum consistent with the GZK prediction \citep{HRspec08,augerSpec08}.  Nonetheless, the puzzle of which astrophysical sites are capable of accelerating UHECRs remains open, and the answer will be of fundamental importance for our understanding of gamma ray bursts (GRBs), active galactic nuclei (AGNs) and quasars, and other extreme systems.  Efforts to find angular correlations between UHECR arrival directions and candidate astrophysical sources have been bedeviled until recently by a combination of inadequate statistics and the fact that UHECRs, being charged particles, are deflected by magnetic fields en route from their sources.  

The Pierre Auger Observatory's discovery of a significant correlation between the highest energy cosmic rays and nearby galaxies in the \citet{VCV} (VCV) Catalog of Quasars and Active Galactic Nuclei (12$^{\rm th}$ Ed.) \citep{augerScience07, AugerLongAGN} (Auger07a,b below) is an important step toward identifying the sources of UHECRs.  Of the 27 cosmic rays above 57 EeV recorded prior to Aug. 31, 2007, twenty are within 3.2$^\circ$ of a VCV galaxy with $z \leq 0.018$ (about 75 Mpc).  Restricting to $|b|>10^\circ$, where the VCV catalog is more complete, there are 22 UHECRs of which 19 are correlated. About 5 of these correlations would be expected by chance if the arrival directions of UHECRs were isotropic.  The distribution of magnetic deflections is not known {\em a priori} and it is not yet known with certainty whether the UHECRs are protons or nuclei. Consequently the expected correlation between UHECR energy and maximum source distance is not certain.  The strategy used by Auger was to search for correlations by scanning over UHECR energy threshold, maximum angular separation, and maximum source redshift, to find the values of these parameters that maximize the significance of the correlations.  The VCV catalog and the parameters given above were identified using data through May 31, 2006 and a ``prescription" formulated for an independent test.  The correlation was confirmed with independent subsequent data, taken from June 1, 2006 to Aug. 31, 2007, with a probability of less than 1\% of occurring by chance.  

This Auger result is of fundamental importance to particle astrophysics, because the correlation with nearby extragalactic structure clearly demonstrates that UHECRs are of extragalactic origin and that the highest energy cosmic rays have a horizon, consistent with the GZK effect. (The downturn in the spectrum might merely be due to a maximum energy of the accelerators.)  However as stressed by the Auger collaboration, the observed correlation may not mean that the correlated UHECRs are produced by galaxies with which they are correlated:  the VCV galaxies may just be tracers of the true sources.   

Our purpose here is to examine the VCV galaxies that correlate with the 20 Auger UHECRs, as a step toward elucidating properties of the sources of UHECRs.  We emphasize that the correlation observed by Auger is only statistical: $\sim$1/4 of the correlations are expected simply by chance.  We cannot be confident that any given one of the correlated VCV galaxies is a source.  However the degree of correlation observed by Auger is so high that galaxy clustering alone cannot account for it (GRF, A. Berlind and IZ, in preparation); of order half or more of the correlated galaxies are most likely the sources of their associated UHECR, enabling us to obtain statistically useful information on source properties even though we do not have a pure sample of sources.  The underlying principle is that the chance correlations must be representative of the ensemble of VCV galaxies, and therefore their presence does not distort the conclusions when properly used.

The VCV catalog is a list of the active galactic nucleus, quasar and BL Lac candidates reported in the literature, based on heterogeneous selection criteria.  While it is the largest available collection of known AGNs, especially for the southern hemisphere where most of Auger's exposure lies, it has several deficiencies compared to an ideal catalog. The VCV catalog is known to be incomplete and non-uniform. Furthermore, it is not pure.  Here we assemble the correct classifications of the VCV galaxies correlated with the Auger UHECRs.  In addition, we determine the bolometric luminosity of each AGN, which is an important diagnostic of the maximum UHECR energy an AGN can produce in conventional jet acceleration.  Finally, we look for evidence of a bolometric luminosity threshold for AGNs which may be responsible for UHECR production.  We examine the AGNs in the Virgo region, from which no UHECRs were detected, and find that most are much lower in their bolometric luminosity than the AGNs which are correlated with UHECRs.  Thus the lack of UHECRs from the direction of Virgo may simply reflect the existence of a minimum bolometric luminosity of AGNs responsible for accelerating UHECRs.  We also estimate the fraction of low-luminosity AGNs (LLAGNs, taken here to have $L_{\rm bol} \leq 5 \times 10^{42} {\rm erg\, s^{-1}}$) in the VCV catalog in the HiRes exposure region, finding that about half the AGNs are below the luminosity threshold of AGNs correlated with Auger UHECRs.  The fraction of known LLAGNs in VCV is higher in the HiRes field of view than in Auger's, due to the sensitive northern hemisphere survey by \citet{Ho1995}; this may contribute to HiRes' not observing  a significant correlation between VCV galaxies and their UHECR data \citep{HRAGN08}.

\section{UHECR acceleration}
\label{accel}
The motivation and agenda for our study derive from theoretical considerations of UHECR acceleration.   No astrophysical system has been conclusively demonstrated theoretically to be capable of accelerating cosmic rays to the observed energies of $\gtrsim 10^{20} $eV.  GRBs have been argued to be responsible for UHECRs \citep{waxman95,waxmanUHECR} on the basis of their energy injection rate and theoretical plausibility:  they are known to produce high energy photons and the GRB internal-shock model can be viable for UHECR acceleration as well. Other possible accelerators include internal shocks in the jets of Active Galactic Nuclei -- analogous to those in GRBs but with much lower bulk Lorentz factors (for an early suggestion see \citet{biermannStrittmatter87}), external shocks in the lobes of powerful radio galaxies, and magnetars \citep{aronsUHECR}, to name some of the more popular.  A GRB could satisfy the requirements of UHECR acceleration for $\Gamma \sim 10^3$ \citep{waxman95}, while for AGN jet acceleration $\Gamma \sim$ few would be envisaged.  Assuming the correlation between UHECRs and VCV galaxies observed by Auger is real, AGNs become the favored sources and we focus here on testing AGN-based acceleration models.  

A very general requirement to accelerate a cosmic ray proton of energy $E \equiv E_{20} \,10^{20}$ eV, in a relativistic jet of bulk Lorentz factor $\Gamma$, is that the Poynting luminosity of the jet, $L_{\rm P}$, satisfy $  L_{\rm P} \gsi 10^{45} \, \Gamma ^2\, E_{20}^2 \, {\rm erg \, s^{-1}}$ \citep{FarrarandGruzinov}.  This follows because the Larmour radius of the UHECR\footnote{Assumed here to be a proton.}, 0.1 Mpc $E_{20}/B_{\mu \rm G}$, must be less than the characteristic size of the acceleration region, $R$, leading to $RB\gsi 3\times 10^{17}\,\Gamma ^{-1} \, E_{20} \,{\rm G \, cm}$, which translates to a lower bound on the Poynting luminosity of the jet, $L_{\rm P} \sim {1\over 6}c\Gamma ^4B^2R^2$.  For a conventional AGN jet the Poynting luminosity is supplied by accretion, so both the jet and accretion disk of the active nucleus must satisfy the quoted luminosity constraint, as discussed in greater detail by \citet{fg08}.   Therefore, the bolometric luminosity of the accretion disk of an AGN capable of accelerating a proton to $E_{20} \,10^{20}$ eV must satisfy
\be
\label{Lbolmin}
 L_{\rm bol} \gsi 10^{45} \, \Gamma ^2\, E_{20}^2 \, {\rm erg \, s^{-1}}.
 \ee  
If the time-scale of variation of the accelerator is large compared to the cosmic ray travel time, estimated to be $\lesssim 10^5$ yr for the correlated Auger cosmic rays \citep{fg08}, then the required luminosity should be evident when observing the sources today.  In conventional AGN jet acceleration or acceleration in radio lobes, the time scale for variation of the source is of order the lifetime of an AGN, i.e., $\gtrsim 10^7$ yr.  Thus an AGN capable of accelerating UHECRs should either have powerful radio lobes or have an accretion disk luminosity consistent with the bound on $L_{\rm bol}$ in equation (\ref{Lbolmin}).  

An alternative to the conventional continuous jet model of AGN acceleration has recently been proposed by \citet{fg08}.  In this ``Giant AGN Flare" mechanism, an instability of the accretion disk -- perhaps initiated by the tidal disruption of a passing star -- produces an intense flare lasting of order a day to a month.   During the flare the luminosity condition for UHECR acceleration is easily satisfied \citep{fg08}, but afterward the emission subsides quickly because the cooling time is short and the material which fuels the flare is largely consumed, so the system as observed today need not satisfy the luminosity condition (\ref{Lbolmin}).   In section \ref{sec:Lbol} we determine, for each AGN-UHECR pair, the figure-of-merit $ \lambda_{\rm bol} \equiv L_{\rm bol} 10^{-45} E_{20}^{-2}$.  $ \lambda_{\rm bol} $ should be $\gsi 1$ for conventional continuous jet acceleration of protons, but has no such constraint for the Giant Flare scenario.   We find that at least half of the correlated AGNs do not satisfy the bolometric luminosity requirement, favoring the Giant AGN Flare scenario.

A question raised (but not yet answered) by the Giant AGN flare model, is whether the AGNs responsible for UHECR production, as observed today, show a threshold $L_{\rm bol}$.  In the AGN flare model an accretion disk is required {\em prior} to the flare, but the observed value of $L_{\rm bol}$ reflects the properties of the {\em remnant} system, which depends on how much material is left after the period of rapid accretion, and how cool the accretion disk has become in the intervening time.  In section \ref{sec:thresh} we find empirical evidence of a threshold $L_{\rm bol} \geq 5 \times 10^{42} {\rm erg \, s^{-1}}$.

We now turn to our task; our first objective is to identify the AGNs among the correlated VCV galaxies.

\section{Identification of Active Galactic Nuclei}

There are a number of diagnostics used to identify an accreting
supermassive Black Hole from optical spectroscopy.  Broad permitted emission lines
(e.g. H$\alpha$) with line widths $\gtrsim$ 1000 km/s strongly suggest
that we are seeing the Keplerian velocities of gas orbiting the BH.  This
dense orbiting gas, known as the broad-line region, has a typical size of
light days, and thus is never spatially resolved in astronomical
observations. Objects with observable broad emission lines, for which we
have an unobstructed view of the nucleus, are known as broad-line AGNs
(or, at low luminosities, Seyfert 1 galaxies). In some cases our view of
the central engine and broad-line region are obscured.  AGNs may still be
identified by the presence of narrow emission lines, originating from more
rarefied gas at scales of 100-1000s of pc.  In particular, the ratios of
strong emission lines (typically H$\beta$, [\ion{O}{3}]~$\lambda 5007$,
H$\alpha$, and [\ion{N}{2}]~$\lambda 6584$) are very sensitive to the
shape of the ionizing continuum, and thus allow us to distinguish between
nebulae excited by starlight, shocks, or the hard continuum of a radiating
BH.  Obscured objects are known as narrow-line AGNs (Seyfert 2 galaxies at
low luminosities).

Two-dimensional line ratio diagnostics have been developed to effectively
discriminate between various emission mechanisms, e.g., star formation,
shock, or photoionization by an accreting BH.  See \citet{BPT1981}(BPT below), \citep{VO1987,
IC5169, Kauffmann2003}.  Baldwin, Phillips, \& Terlevic (BPT) diagnostic
diagrams are based on the relative strengths of prominent emission lines 
(e.g.~[\ion{O}{3}]/H$\beta$ versus [\ion{N}{2}]/H$\alpha$) that are close
together in wavelength space, to minimize the impact of reddening.  A
typical BPT diagram is shown in Figure \ref{BPT}, along with various
boundary lines between different types of galaxies with narrow-line
emission.  \citet{IC5169} developed a conservative boundary between star
formation and AGN activity based on theoretical modeling, shown as a solid
curve in Figure \ref{BPT}; galaxies lying above and to the right of the
Kewley line are unambiguously AGNs.  A looser, empirical boundary was used
by \citet{Kauffmann2003} to identify SDSS star-forming galaxies by their
line ratios; it is shown as a dashed curve in Figure \ref{BPT}.  There is
a further distinction for the objects which fall outside the star
formation boundary: those with log([\ion{O}{3}]/H$\beta$) $>$ 0.48, are
known as high-ionization Seyfert galaxies, and those with
log([\ion{O}{3}]/H$\beta$) $\leq$ 0.48 are known as low-ionization nuclear
emission regions (LINERs) \citep{Heckman1980}.  The Seyfert-LINER boundary
is indicated by the horizontal line in Figure \ref{BPT}.  While many
LINERs have been demonstrated to be powered by accretion activity (see
review in \citet{Ho2008}), there are other processes such as shocks
\citep{Veilleux} which can lead to LINER-like line ratios.  Caution, and
auxiliary data, are often required to determine the nature of distant LINERs.

It is, of course, important to remember that all AGN selection techniques
are biased in different ways, and are dependent on the depths and
apertures of individual surveys.  For nearby AGNs, optical
emission line surveys are probably the most complete, e.g., as shown by \citep{Heckman&2005,Ho2008}.  Note, however, that \citet{reviglioHelfand06} find that about half the AGNs
detected in their sample using radio or X-ray selection are not identified by the BPT
criteria discussed above.  Since UHECR source candidates are closer than
about 100 Mpc, and star-forming regions with enough emission to be
confused with an AGN are larger than $\approx 10$ pc, X-ray observation
with Chandra resolution can decide whether a UHECR candidate source is an
AGN or not.  In the radio, detection of jets unambiguously identifies an
AGN. If no jets are seen, the presence of an AGN can be inferred if there
is a compact source at the center of the galaxy whose radio emission
exceeds that which would be expected for a nuclear starburst region based
on the far infrared-radio relation \citep{Condon1992}.

\section{UHECR-correlated VCV Galaxies}
\label{sec:UHECR-corr}
\begin{figure}[h]
\begin{center}
\includegraphics[scale=0.6]{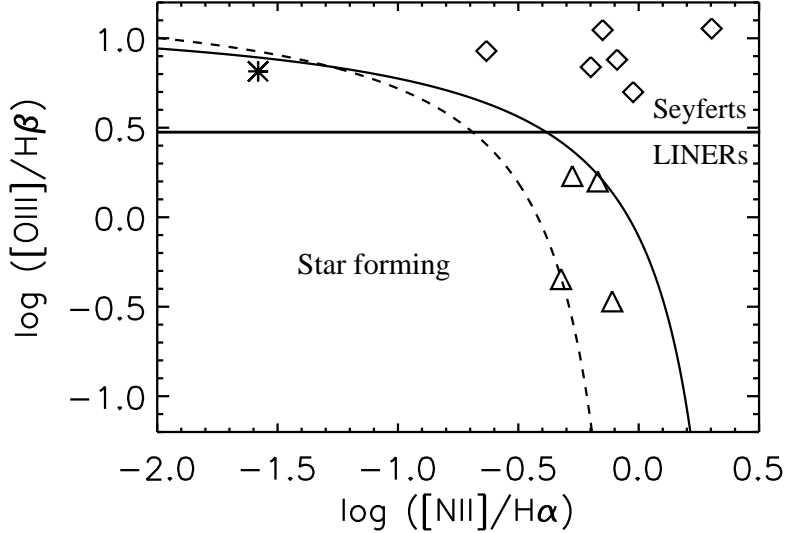}
\end{center}
\vspace{-0.5cm}
\caption{A BPT diagram 
showing the line ratios, [OIII]/H$\beta$ vs. [NII]/H$\alpha$, for 11 of the 15 VCV galaxies correlated with UHECRs which are not a broad-line AGN or BL Lac.  NGC 5244 and NGC 7135 are not plotted since they show almost no [OIII] emission, nor are NGC 4945 and ESO 139-G12 since their identification as AGNs is made on other grounds. The dwarf galaxy, Q 2207+0122, is indicated by the asterisk. NGC 2989, IC 5169, NGC 1204, and NGC 7591 are shown as triangles.  These fall to the left and below the theoretical classification line for AGN developed by \citet{IC5169}, indicated by the solid line.  The Seyfert 2 galaxies are shown as diamonds. They lie well within the Seyfert region of the diagram. The demarcation line used by \citet{Kauffmann2003} is shown (dashed) for comparison. The horizontal line shows the boundary between Seyferts (above the line) and LINERs (below the line).  
\label{BPT}}
\vspace{-0.75cm}
\end{figure}

Auger has published 27 UHECRs with energies above 57 EeV, 20 of which are correlated with a VCV galaxy closer than $z = 0.018$ (Auger07a,b).  That is our basic UHECR sample for this study.  The 20 UHECRs are correlated with 21 VCV galaxies.\footnote{Some VCV galaxies have 2 UHECRs within $3.2^\circ$ and some UHECRs have multiple VCV galaxies within $3.2^\circ$.}  Table \ref{matchedAGN1} shows the 20 correlated UHECRs and the 21 associated VCV galaxies.  For each VCV galaxy, we check the VCV classification against literature and by examining available spectra.   Five are unambiguously broad-line AGNs and one is a BL Lac -- widely considered to be a radio-loud AGN viewed directly down the jet.  The remaining 15 galaxies are potential narrow-line AGNs and have to be evaluated individually.  Figure \ref{BPT} is a BPT diagnostic diagram in which the line ratios, log([OIII]/H$\beta$) vs. log([NII]/H$\alpha$), for 11 of the 15 galaxies are plotted.  The four galaxies which cannot be plotted in Fig. \ref{BPT} consist of two AGNs and two without nuclear activity. The optical emission from NGC 4945 is highly obscured but its hard X-ray spectrum clearly identifies AGN activity \citep{Madejski2000}. ESO 139-G12 shows a hint of a broadened H$\alpha$ line \citep{Marquez2004} and is an AGN. NGC 5244 \citep{NGC5244} and NGC 7135 \citep{NGC7135} have almost no [OIII] emission; they cannot be identified as AGNs from their optical spectra. 

Eleven galaxies appear in Figure \ref{BPT} of which six, shown as diamonds on the BPT diagram, fall comfortably in the Seyfert region of the BPT diagram; these are clear AGNs.  Two of the eleven (NGC 7591 and Q 2207+0122), shown as a triangle and an asterisk, fall below the \citet{Kauffmann2003} empirical star formation line and are not AGNs. The remaining three (NGC 2989, NGC 1204, IC 5169), shown as triangles, fall between the Kewley and Kauffmann lines, in the LINER region of the BPT diagram.  \citet{IC5169} uses additional information from other emission lines and determines IC 5169 to be purely star forming;  NGC 2989 \citep{Phillips1983} and NGC 1204 \citep{Corbett2003} are predominantly star-forming galaxies but may have limited AGN activity, contributing $\leq$ 25\% of the total luminosity.  Detailed information on each of the correlated VCV galaxies is given in Appendix \ref{section:individualobjects} and summarized in Tables \ref{Lbol} and \ref{LIR}.  

In all, 14 of the 21 VCV galaxies, correlated with 14 UHECRs, are AGNs, while seven show limited or no sign of AGN activity.  Radio or X-ray observations of these 7 VCV galaxies which do not appear to be AGN are needed to determine whether  or not all UHECRs may be attributable to AGNs.\footnote{A proposal has been submitted to Chandra by GRF, W. Terrano and IZ to make the observations required.} Figure \ref{skyplot} shows Auger's 27 highest-energy cosmic rays with the correlated UHECRs represented by 3.2$^\circ$ circles and the uncorrelated ones represented by 8-point stars. The 14 confirmed AGNs are shown as triangles and the 7 other correlating VCV galaxies are shown as squares.  Auger's exposure cutoff at $\delta$ $=$ $25^\circ$ is indicated (cyan line) as is the Galactic plane band ($|b| = \pm10^{\circ}$), where VCV is especially incomplete.  The rest of the VCV galaxies with z $\leq$ 0.018 are shown as points, color-coded in bins of redshift.  

\begin{table}
\begin{center}
\caption{Auger UHECRs and Correlated VCV Galaxies \label{matchedAGN1}}
\begin{tabular}{cccccccccccccc}
\hline
\hline
Year & Day & E$_{CR}$ & RA$_{CR}$ & Dec$_{CR}$ & $l_{CR}$ & $b_{CR}$ & VCV Galaxy &  RA$_{AGN}$ & Dec$_{AGN}$ & r &  z  &   VCVClass &  RealClass \\
& & (EeV) & (J2000) & (J2000) & (J2000) & (J2000) & & (J2000) & (J2000) & (deg) & & & \\
\hline  
$*$2004 & 142  &    84   &    199.7 &  $-$34.9  & 309.2 & $+$27.6 & ESO 383-G18  &  203.36  &  $-$34.02 & 3.14 & 0.013  &   S1.8  &  S2   \\
$*$2004 & 282  &    66   &    208.0 &  $-$60.3  & 310.4 & $+$1.7 & 4U 1344-60  &  206.90  &  $-$60.62 & 0.63 & 0.013  &   S1    &  S1   \\
$*$2004 & 339  &    83   &    268.5 &  $-$61.0  & 332.3 & $-$17.0 & ESO 139-G12  &  264.41  &  $-$59.94 & 2.27 & 0.017  &   S2    &  S2   \\
$*$2004 & 343  &    63   &    224.5 &  $-$44.2  & 325.6 & $+$13.0 & IC 4518A    &  224.43  &  $-$43.13 & 1.07 & 0.016  &   S2    &  S2   \\
$*$2005 &  54  &    84   &     17.4 &  $-$37.9  & 284.4 & $-$78.6 & NGC 424     &   17.87  &  $-$38.08 & 0.41 & 0.011  &   S1h   &  S1.5 \\
2005    &  63  &    71   &    331.2 &   $-$1.2  &  58.8 & $-$42.4 & Q 2207+0122   &  332.62  &    $+$1.62 & 3.16 & 0.013  & $\cdots$  &  HII$^1$ \\ 
$*$2005 &  81  &    58   &    199.1 &  $-$48.6  & 307.2 & $+$14.1  &  NGC 4945    &  196.37  &  $-$49.47 & 1.99 & 0.002  &    S    &  S2    \\
2005    & 295  &    57   &    332.9 &  $-$38.2  &   4.2 & $-$54.9 & IC 5169     &  332.54  &  $-$36.09 & 2.13 & 0.010  &   S2       &  HII$^2$ \\
$*$2005 & 306  &    59   &    315.3 &   $-$0.3  & 48.8 & $-$28.7 & Zw 374.029   &  313.84  &  $+$2.35  & 3.03 & 0.013  &   S1h   &  NLS1  \\
$*$2006 &  35  &    85   &     53.6 &   $-$7.8  & 194.1 & $-$46.9 & SDSS J03302-0532 &  52.56 &   $-$5.54 &  2.48 & 0.013 &  S1    &  S1    \\
     &      &         &          &         &  & &  NGC 1358    &   53.42   &  $-$5.09  & 2.72 & 0.013 &   S2       &  S2    \\
     &      &         &          &         &  & &  SDSS J03349-0548 &  53.74 &   $-$5.81 & 1.99 & 0.018 &   S1       &  S1    \\
$*$2006 &  55  &  59   &    267.7 &  $-$60.7    & 332.4 & $-$16.5 &  ESO 139-G12  &  264.41   &  $-$59.94 & 1.79 & 0.017 &   S2    &  S2    \\
2006 & 185  &    83   &    350.0 &   $+$9.6  & 88.8 & $-$47.1 &  NGC 7591    &  349.57  &  $+$6.59   & 3.04 & 0.017 &   S        &  HII$^3$ \\ 
$*$2006 & 296  &    69   &     52.8 &   $-$4.5  & 189.4 & $-$45.7 & NGC 1358    &   53.42  &  $-$5.09   & 0.85 & 0.013 &   S2    &  S2    \\
     &      &         &          &        & & & SDSS J03302-0532 &  52.56 &   $-$5.54  & 1.07 & 0.013 &   S1       &  S1    \\ 
     &      &         &          &         & & &  MARK 607    &   51.20  &  $-$3.42   & 2.16 & 0.009 &   S2       &  S2    \\
     &      &         &          &         & & & SDSS J03349-0548 &  53.74 &   $-$5.81 & 1.61 & 0.018 &   S1       &  S1    \\
$*$2006 & 299  &    69   &    200.9 &  $-$45.3 & 308.8 & $+$17.2 & NGC 5128 (Cen A) & 201.37 &  $-$43.02  & 2.31 & 0.001 &   ?     &  BL Lac  \\
     &      &         &          &         & & &  NGC 5244      & 204.66 &  $-$45.86  & 2.70 & 0.008 &   H2       &  HII$^4$ \\ 
2007 &  51  &    58   &    331.7 &   $+$2.9  & 63.5& $-$40.2 & Q 2207+0122    & 332.62 &   $+$1.62  & 1.58 & 0.013 & $\cdots$   &  HII$^1$ \\ 
$*$2007 &  69  &    70   &    200.2 &  $-$43.4 & 308.6 & $+$19.2 & NGC 5128 (Cen A) & 201.37 &  $-$43.02  & 0.93 & 0.001 &   ?     &  BL Lac   \\
2007 &  84  &    64   &    143.2 &  $-$18.3  & 250.6 & $+$23.8 &  NGC 2989   &   146.36  &  $-$18.37  & 3.00 & 0.013 &   H2       &  HII$^5$  \\
2007 & 145  &    78   &     47.7 &  $-$12.8  & 196.2& $-$54.4 &  NGC 1204   &    46.17  &  $-$12.34  & 1.57 & 0.014 &   S2       &  HII$^6$  \\
$*$2007 & 193  &    90   &    325.5 &  $-$33.5  & 12.1 & $-$49.0 &   IC 5135   &   327.08  &  $-$34.95  & 1.95 & 0.016 &   S1.9  &  S2     \\
     &      &         &          &         & & & NGC 7135    &  327.44   & $-$34.88  & 2.12  & 0.007 & $\cdots$   &  HII$^7$  \\        
$*$2007 & 221  &    71   &    212.7 &   $-$3.3  & 338.2 & $+$54.1 &  NGC 5506   &   213.31  &  $-$3.21  & 0.62  & 0.007 &   S1i   &  S2     \\
\hline
\hline
\end{tabular}
\tablecomments{This table lists each Auger UHECR (with E $>$ 57 EeV) whose arrival directions are within 3.2$^{\circ}$ of a nearby (z $\leq$ 0.018) VCV galaxy, along with the correlated VCV object. The year and the Julian day when the UHECRs were recorded are given, as well their energies and positions (equatorial and galactic) in degrees, from Auger07b. This is followed by the name of the VCV galaxy, its position in degrees, and its separation from the UHECR in degrees, its redshift, and the VCV classification.\footnote{Star forming regions are also known as HII or H2 regions because young stars ionize the hydrogen around them (neutral hydrogen is known as HI).}  The last column shows the correct classification (taken from the literature where available and confirmed in every case). The UHECRs correlated with galaxies that are shown here to have active nuclei are marked with an asterisk.\\} 
\tablerefs{(1) \citet{Q2207}, (2) \citet{IC5169}, (3) \citet{Sturm}, \citet{Veilleux}, (4) \citet{NGC5244}, (5) \citet{NGC2989}, (6) \citet{Sturm}, \citet{Veilleux}, (7) \citet{NGC7135}.}
\end{center}
%\vspace{-0.75cm}
\end{table}

%-----------------------------------------------------------------------
%-----------------------------------------------------------------------

%-----------------------------------------------------------------------
\begin{figure}[h]
%\epsscale{1.0}
%\vspace{-0.5cm}
\begin{center}
\end{center}
%\vspace{-2.0cm}
\caption{Sky plot in equatorial coordinates showing Auger's 27 highest energy cosmic rays and VCV galaxies. The UHECRs correlating with an AGN are indicated with open $3.2^\circ$ circles and those that do not are indicated with an 8-point star. The 14 correlated VCV galaxies which are known to be AGNs are shown as triangles.  The 7 other correlating VCV galaxies are shown as squares.  The rest of the VCV galaxies with $z \leq 0.018$ are shown as points color-coded by redshift. The Galactic plane band ($|b| = 10^{\circ}$) where the VCV catalog is most incomplete is indicated by the solid black lines. Auger's exposure cutoff at $\delta$ $<$ $25^\circ$ is indicated by the cyan line. \label{skyplot}}
\end{figure}
%-----------------------------------------------------------------------

\section{Bolometric Luminosities}
\label{sec:Lbol}

Under the widely-used assumption of a universal spectral energy distribution for active galaxies, the bolometric luminosities may be estimated from measurements of a single line or in a single band.  In most cases, we have flux measurements in only one waveband.  In this paper, we use the following bolometric corrections for [OIII] and broad H$\alpha$ emission: $L_{\rm bol} = 3500~L_{\rm [OIII]}$, with a 0.38 dex variance, as determined by \citet{OIIIconv}, and $L_{bol} = (L(H\alpha_{broad})/10^{42})^{0.86} \times 2.34 \times 10^{44} \rm{erg s^{-1}}$ \citep{GreeneandHo2007}.  

The conversion factor from the 2-10 keV X-ray luminosity to the bolometric luminosity, as determined by  \citet{LxConv}, is dependent upon the level of AGN activity; it is 15-25 for AGNs with Eddington ratios ($L_{bol}/L_{Edd}$) below $\sim$0.1, and 40-70 for those with Eddington ratios above $\sim$0.1. The Eddington luminosity, $L_{Edd}$, is related to the SMBH mass, M, by the relation $L_{Edd} = 3.3 \times 10^4 (M/M_{\odot}) L_{\odot}$. We were able to obtain SMBH mass estimates for four AGNs in our sample; two can be found in the literature and we derive two from the broad H$\alpha$ luminosity and line width following \citet{GreeneandHo2005}. (See details in Appendix A.) All four have low Eddington ratios: 0.006, 0.015, 0.029, and 0.1.   Therefore, in the absence of SMBH mass measurements for the rest, we use $L_{\rm bol} = 20 L_{2-10 \rm keV}$ for all.

We determine the bolometric luminosities for the correlated AGNs from [OIII], broad H$\alpha$, or 2-10 keV emission.   For the broad line AGNs, the [OIII] and H$\alpha$ luminosities are taken from published data when possible, or fit from available optical spectra. (See Appendix A for details.)  We use multicomponent Gaussian models to fit for the broad H$\alpha$ luminosities. When fitting for [OIII] luminosities, we assume only a single Gaussian. The 2-10 keV X-ray luminosities are collected from literature. There is one AGN (ESO 139-G12), for which neither a calibrated spectrum nor an X-ray detection is available, and we use the upper limit for the X-ray luminosity. The AGNs and their bolometric luminosities are given in Table \ref{Lbol}.  

For each AGN-UHECR pair, the figure-of-merit $ \lambda_{\rm bol} \equiv L_{\rm bol} 10^{-45} E_{20}^{-2}$ is determined. Out of the 13 UHECRs which are correlated with AGNs for which we were able to determine the bolometric luminosities, we see that only one UHECR correlates with an AGN having a value of $\lambda_{\rm bol} \gtrsim 1$.  Another 5 UHECRs are correlated with AGNs whose luminosity is within a factor of 2 of the minimum.  Given the rough nature of the $\lambda_{\rm bol}$ bound, these AGN may have sufficient power to accelerate their associated UHECRs.  However seven correlated UHECRs are associated with AGNs that fall far short of the minimum power.  This poses a serious problem for conventional models of UHECR acceleration in continuous AGNs, although it is compatible with the recently proposed giant AGN flare scenario of \citet{FarrarandGruzinov}.

\begin{table}
\label{AGN}
\begin{center} 
\caption{Bolometric Luminosities of Correlated AGN \label{Lbol}}
\begin{tabular}{ccccccccc}
\hline
\hline
AGN &   AugerYr,Day &  E &  	 r  &     z  &   Type & log($L_{\rm IR}$)   & $L_{bol}$  &  $\lambda_{\rm bol}$ \\ 
    &                & (EeV) & (deg) &        &        & ($L_{\odot}$)  & (erg s$^{-1}$) & \\
\hline
NGC 5128(Cen A) & 2006,299  &    69   &   2.31 &  0.001  &  BLL& 10.14 & (0.7-1.4) $\times$ 10$^{43}$  &  0.02-0.03 \\
	        & 2007,69   &    70   &   0.93 &         &     &       &                   &  \\
SDSSJ03302-0532 & 2006,35   &    85   &   2.48 &  0.013  &  S1 & 10.54 & $2.31 \times 10^{43}$  &  0.03 \\
                & 2006,296  &    69   &   1.07 &         &     &       &                        &  0.05 \\ 
SDSSJ03349-0548 & 2006,35   &    85   &   1.99 &  0.018  &  S1 & --- & $5.44 \times 10^{42}$    &  0.008 \\
                & 2006,296  &    69   &   1.61 &         &     &       &                        &  0.01  \\ 
Zw 374.029   &    2005,306  &    59   &   3.03 &  0.013  & NLS1& --- & $2.96 \times 10^{43}$ & 0.085\\
NGC 424   &       2005,54   &    84   &   0.41 &  0.011  &  S1 & 10.61 & $3.73 \times 10^{44}$  &  0.53 \\
4U 1344-60   &    2004,282  &    66   &   0.63 &  0.013  & S1.5& --- & $2.70 - 2.81 \times 10^{44}$ & 0.62 - 0.65 \\
NGC 5506     &    2007,221  &    71   &   0.62 &  0.007  &  S2 & 10.41 & $4.61 - 15.5 \times 10^{43}$ & 0.09 - 0.31\\  
IC 5135      &    2007,193  &    90   &   1.95 &  0.016  &  S2 & 11.35 & $1.23 \times 10^{46}$ & 15.2 \\
MRK 607   &       2006,296  &    69   &   2.16 &  0.009  &  S2 & 10.15 & $3.89 \times 10^{44}$  &  0.82 \\     
IC 4518A     &    2004,343  &    63   &   1.07 &  0.016  &  S2 & --- & $2.44 \times 10^{44}$  & 0.61 \\
NGC 1358  &       2006,296  &    69   &   0.85 &  0.013  &  S2 & --- & $2.51 \times 10^{44}$  &  0.53 \\
                & 2006,35   &    85   &   2.72 &         &     &       &                       & 0.35 \\ 
ESO 383-G18  &    2004,142  &    84   &   3.14 &  0.013  &  S2 & 10.48 & $4.62 \times 10^{43}$ & 0.065\\
NGC 4945  &       2005,81   &    58   &   1.99 &  0.002  &  S2 & 10.46 & $2 \times 10^{44}$  &  0.06 \\ 
ESO 139-G12  &    2004,339  &    83   &   2.27 &  0.017  &  S2 & 10.22 & $< 9.5 \times 10^{44}$ & $<$ 1.4 \\
	     &    2006,55   &    59   &   1.79 &         &     &       &                   &  $<$ 2.7 \\
\hline
\hline
\end{tabular}
\tablecomments{This table lists key properties of the VCV AGNs which are within 3.2$^\circ$ of an Auger UHECR. The year, Julian day, and energy (in EeV) of the correlated UHECR are given for each AGN. The separation between the AGN and UHECR are given in degrees. The redshift and type of the AGN are given. (NLS1 indicates the presence of both narrow and broad line emission.) The derived infrared and bolometric luminosities are listed. $\lambda_{\rm bol}$ is defined as $L_{bol} \times 10^{-45} \times \rm{E_{20}}^{-2}$, where E$_{20} \equiv E/100$; it is the ratio which compares the bolometric luminosity to the theoretical luminosity necessary for producing the associated UHECR. }
\end{center}
%\vspace{-0.5cm}
\end{table}

\section{Possible Threshold in Bolometric Luminosity}
\label{sec:thresh}
%\section{AGNs in the Virgo Region}
We have seen in section \ref{sec:Lbol} that only one of the AGNs correlated with UHECRs satisfies the naive limit for UHECR acceleration.  We seek in this section to determine whether correlated AGNs show any threshold bolometric luminosity -- such a threshold would elucidate the nature of the UHECR acceleration mechanism, and as we shall argue, can account for puzzling features of present observations.

One puzzling aspect of the observed correlation of Auger UHECRs with nearby VCV galaxies is the lack of UHECRs from the Virgo Cluster (Auger07b).  Because of its proximity, and the very large number of VCV galaxies in Virgo, this seems quite surprising.  Therefore, we have made the same analysis as described above, for VCV galaxies in Virgo.  We examined the VCV galaxies within a circle of radius 15$^{\circ}$ centered on the Virgo Cluster at $\alpha_{\rm J2000}$ = 12:26:32.1, $\delta_{\rm J2000}$ = +12:43:24 \citep{Virgo} and z $\leq$ 0.009, which are attributed to the Virgo Cluster by (or are not listed in) \citet{Ho1997a}. There are a total of 30 VCV galaxies which satisfy these criteria. VCV classifies five of them as HII galaxies. One other galaxy was also found to be an HII galaxy. We can determine the bolometric luminosities of 23 out of the remaining 24.  In 18 cases we use the measured 2-10 keV X-ray luminosities \citep{Ho2001, Terashima2002, Satyapal2005, Gonzalez2006, Panessa2006, Horst2008, Kandalyan2005}, and for the five for which X-ray luminosities are not available we use the measured [OIII] flux \citep{Shields2007, Ho1997a}. There is some evidence that the [OIII] to bolometric luminosity conversion depends on luminosity or Eddington ratio (e.g., \citet{Netzer2006}).  Since optical emission line fluxes, particularly H$\alpha$, may be contaminated by non-nuclear sources (e.g., \citet{Ho2001}) we have chosen to use the hard X-ray flux as a bolometric indicator whenever possible. 

\begin{figure}[h!]
%\epsscale{1.0}
\begin{center}
\includegraphics[scale=0.5]{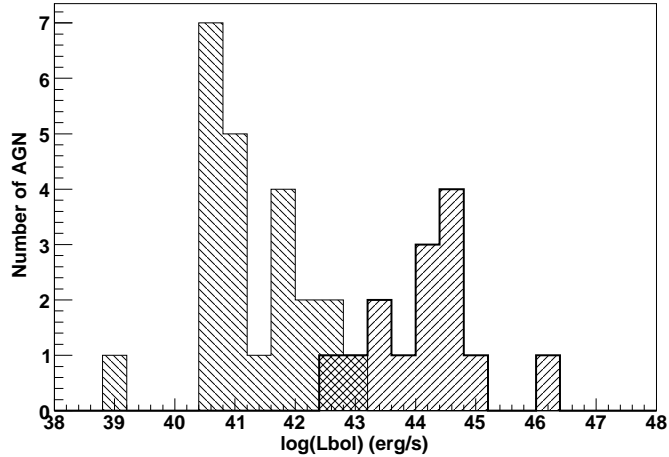}
\end{center}
\vspace{-0.5cm}
\caption{Histogram of the bolometric luminosities of Virgo AGNs (thin outline) compared with the bolometric luminosities of the AGNs correlated with UHECRs (thick outline). The correlated AGNs are much more luminous than the Virgo AGNs on average. \label{Virgo}}
\end{figure}

The bolometric luminosities of the Virgo AGNs, listed in Appendix \ref{section:VirgoDetails}, are systematically lower than those of the AGNs correlated with UHECRs (which span the range 5 $\times$ 10$^{42}$ to 1 $\times$ 10$^{46}$ erg s$^{-1}$). Of the Virgo AGNs, only two have bolometric luminosities in the range of the correlated AGNs (NGC 4338 and NGC 3976, with luminosities of 1.1 $\times$ 10$^{43}$ erg s$^{-1}$ and 4.6 $\times$ 10$^{42}$ respectively) and three others (NGC 4486, NGC 4579, and NGC 4698) have bolometric luminosities within a factor of five lower.  The remaining 18 have bolometric luminosities ranging from 10$^{39}$ erg s$^{-1}$ to 10$^{41}$ erg s$^{-1}$, far below the bolometric luminosities of AGNs correlated with UHECRs. A histogram of the bolometric luminosities of the Virgo AGNs, compared with a histogram of the bolometric luminosities of the AGNs correlated with UHECRs, is shown in Figure \ref{Virgo}. The abundance of known low luminosity AGNs in Virgo is due to the sensitive Palomar spectroscopic survey of nearby galaxies in the northern hemisphere by \citet{Ho1995}, which found AGNs in $\sim$60\% of bulge-dominated galaxies \citep{Ho1997b}.  

It may be that the accretion activity of these very low luminosity AGNs, is too low for them to be sources of UHECRs.   A lower bound on the quiescent accretion rate might be expected if UHECRs are accelerated in intense, month-long, giant AGN flares, triggered when the tidal stream of a star disrupted by a SMBH interacts with an existing accretion disk, as proposed by \citet{FarrarandGruzinov}.  Note that if there is a threshold accretion luminosity, it may be larger than $L_{\rm bol,min}$ of the correlated AGNs, since $\sim 3$ of the correlated AGNs are presumably chance associations, and the AGN luminosity function increases at low luminosities \citep{Ho1995}.

%-----------------------------------------------------------------------
%-----------------------------------------------------------------------

Another puzzle is the lack of significant correlation between UHECRs and VCV galaxies reported by the HiRes Collaboration \citep{HRAGN08}. If UHECRs are preferentially accelerated by AGNs above a threshold bolometric luminosity, then one would expect a reduced degree of correlation significance observed by northern hemisphere experiments such as AGASA and HiRes, compared to what is seen by Auger in the southern hemisphere.  This is because a substantial fraction of the VCV northern hemisphere AGNs are LLAGNs \citet{Ho1995} and 14\% of Auger's exposure and 84\% of HiRes' exposure are from the northern hemisphere.  We can estimate the degree of dilution in the VCV catalog for $z \leq 0.018$ in the HiRes exposure compared to that in the Auger exposure, as follows.  The \citet{Ho1995} survey included 417 emission line galaxies with $z \leq 0.018$.  They report that about 50\% are AGN, with median $L(H\alpha) = 2 \times 10^{39}$ erg s$^{-1}$, which roughly corresponds to our empirical UHECR threshold $L_{\rm bol,min} = 5 \times 10^{42}$ erg s$^{-1}$.  Thus at least $\approx 105$ of the VCV galaxies in the northern hemisphere with $z \leq 0.018$ are LLAGNs.  The exposure weighted average fraction of LLAGNs and HII galaxies compared to total VCV galaxies is $\approx 0.443$ for HiRes, but only $\approx 0.193$ for Auger.  This is a concrete illustration of the systematic differences in completeness and purity, between the portions of the VCV catalog used by the two UHECR experiments.  

\section{Infrared Luminosities}

An interesting question is whether the correlated VCV galaxies which are not optical AGNs have any common properties.  One relevant property could be the total infrared luminosity, which is sensitive to both star formation and nuclear activity.  In fact, there is circumstantial evidence that accretion is accompanied by significant star formation (e.g.,~Sanders \& Mirabel 1996; Kauffmann et al. 2003), although the temporal coincidence of the two is a matter of debate (Ho 2005).  It is certainly theoretically plausible that the large gas accretion episodes leading to AGN activity are accompanied by significant star formation events, and correspondingly elevated dust levels. Sufficiently dense and dusty star formation, such as is seen in ultra-luminous infrared galaxies, may in principle completely obscure AGN activity.

We determine the IR luminosities of the seven VCV galaxies that are not AGNs from {\it IRAS} fluxes following the prescription of \citet{SandersandMirabel1996}:
\begin{equation}
\label{LIR_equation}
F_{\rm IR} = 1.8 \times 10^{-14} [13.48 f_{12} + 5.16 f_{25} + 2.58 f_{60} + f_{100}] ~{\rm W/m^2}, L_{\rm IR} = 4 \pi D_L^2 F_{\rm IR},
\end{equation}
where $D_L$ is the luminosity distance of the galaxy, and $f_{12}$ to $f_{100}$ are the $IRAS$ fluxes.
One of these non-optical-AGN galaxies is a luminous infrared galaxy (LIRG) (high rate of star formation: $11 < {\rm log}[L_{IR}/L_{\odot}] < 12$), four are starburst galaxies  ($10 < {\rm log}[L_{IR}/L_{\odot}] < 11$), one is a quiescent galaxy (${\rm log}[L_{IR}/L_{\odot}] < 10$), and one has very low luminosity -- it is in fact a dwarf galaxy. The infrared luminosities and classifications of these galaxies are listed in Table \ref{LIR}. For completeness, we also calculate the IR luminosities of the AGN and include them in Table \ref{Lbol}.  \citet{SandersandMirabel1996} report that about 15\% of LIRGs are Seyferts and another 35\% are LINERS, whereas for starburst galaxies the LINER fraction is about the same but $\lesssim$5\% are Seyferts.  

\begin{table}
\label{nonAGN}
\begin{center} 
\caption{Correlated VCV Galaxies with Limited AGN Activity \label{LIR}}
\begin{tabular}{ccccccccc}
\hline
\hline
AGN &   AugerYr,Day &  E &  	 r  &     z  & log([NII]/H$\alpha$) & log([OIII]/H$\beta$) & log($L_{\rm IR}$(8-1000 $\mu$m)) & Type \\ 
    &                & (EeV) & (deg) &    &  &    & ($L_{\odot}$) & \\
\hline
NGC 2989 & 2007,84 & 64 & 3.00 & 0.013 & $-$0.28 & 0.23 & 10.48 & Starburst (AGN $\leq$ 20\%) \\
\hline
NGC 7591 & 2006,185 & 83 & 3.04 & 0.017 & $-$0.32 & $-$0.35 & 11.07 & LIRG \\
NGC 1204 & 2007,145 & 78 & 1.57 & 0.014 & $-$0.11 & $-$0.47 & 10.83 & Starburst (AGN $\leq$ 25\%)\\
NGC 5244 & 2006,299 & 69 & 2.70 & 0.008 & --- & --- & 10.41 & Starburst \\
IC 5169  & 2005,295 & 57 & 2.13 & 0.010 & $-$0.17 & 0.20 & 10.36 & Starburst \\
NGC 7135 & 2007,193 & 90 & 2.12 & 0.007 & --- & --- & 9.42 & Quiescent \\
\hline
Q 2207+0122 & 2005,63 & 71 & 3.16 & 0.013 & $-$1.58 & 0.82 &$\cdots$ & Dwarf Galaxy \\
            & 2007,51 & 58 & 1.58 & & & \\
\hline
\hline
\end{tabular}
\tablecomments{This table lists the VCV galaxies within $3.2^\circ$ of an Auger UHECR that do not show evidence in their optical spectra of significant nuclear activity.  The year, Julian day, and energy of the correlated UHECR  as well as the redshift of the galaxy and its separation from the correlated UHECR are given. These are followed by line ratios ([NII]/H$\alpha$ and [OIII]/H$\beta$), fit or gathered from literature. The line ratios for NGC 5244 and NGC 7135 are not 
available but both show almost no [OIII] emission. The IR luminosities (8-1000 $\mu$m) are calculated from the {\it IRAS} fluxes according to the prescription of 
\citet{SandersandMirabel1996}. The final column classifies the galaxies according to their IR luminosities, which are indicators of their star formation rates.
} 
\tablerefs{NGC 2989: \citet{NGC2989}, IC 5169: \citet{IC5169}, NGC 1204: \citet{Veilleux}, NGC 7591: \citet{Moustakas2006}. The line ratios (courtesy of J. Moustakas) for 
Q 2207+0122 are fit from the spectrum (courtesy of J. Gunn and D. Schneider).\\}
\end{center}
\vspace{-0.5cm}
\end{table}

\section{Note Added -- \citet{MSPC} }

While this paper was being reviewed by the Pierre Auger Collaboration Publication Committee (submitted on April 27, approved on May 28, 2008) a paper has appeared by \citet{MSPC} (MSPC below) which takes a complementary approach to ours, examining the radio morphology and luminosity of galaxies that might be the sources of the Auger UHECRs.  MSPC extends the set of candidate sources to include 27 galaxies:  19 of the 21 VCV galaxies considered here, plus 8 more galaxies with $z \leq 0.018$ that are also candidate AGNs according to the NASA Extragalactic Database (NED) but which do not appear in the VCV catalog, plus an additional 27 galaxies from VCV and NED in the redshift range 0.018-0.037.   The 19 VCV $z \leq 0.018$ galaxies that MSPC consider are those in our Table 1, less SDSS J03349-0548 and ESO 383-G18.  The former is removed because they exclude SDSS galaxies (except that SDSS J053302-0532 is in their sample as NGC 1346) but we do not understand why they do not include ESO 383-G18. 

Of the 54 galaxies considered, MSPC reports that 4 of them display the powerful jets or radio lobes expected for UHECR sources: 2 in each of their redshift bins.  One of the two in the lower redshift bin, Cen A (NGC 5128) with $\lambda_{\rm bol}=0.03$, is in our sample and the other, PKS 1343-60, is not.  We examine the optical properties of PKS 1343-60 found in literature. Although it has a bright optical nucleus, the emission is highly absorbed and it could not be determined as an AGN using opical criteria (See Appendix C for details). This most likely explains its absence from the VCV catalog which is largely composed of AGNs identified by optical criteria. We note that while Cen A has the largest radio flux of their candidate sources, 54 Jy, its extreme proximity makes it a weaker radio source than PKS 1343-60 with 23 Jy. 

For completeness, we have extended the study reported in section \ref{sec:UHECR-corr}, to include the 8 new AGN candidates with $z \leq 0.018$ within $3.2^\circ$ of an Auger UHECR introduced by MSPC.  Only two have published optical line fluxes, WKK 4374 (IGR J14515-5542) and NGC 7626.  The former, WKK 4374,  is a narrow line Seyfert (S2) with $L_{\rm bol} = 1.9 \times 10^{44} {\rm erg \, s^{-1}}$ and $\lambda_{\rm bol} = 0.46$; intriguingly, it correlates with a previously uncorrelated UHECR (Auger year, day = 2007, 186; E=64 EeV).  The latter is a radio-loud LINER with $L_{\rm bol} = 1.2 \times 10^{42} {\rm erg \, s^{-1}}$, $\lambda_{\rm bol} = 0.002$.  Two of the UHECRs whose VCV-correlated galaxies were found here to not be identifiable AGNs, (2006, 185; 83 EeV) and (2007, 84; 64 EeV), correlate with the new ``MSPC" source candidates NGC 7626 and NGC 2907 respectively, although a spectrum is needed to determine whether the latter meets optical AGN criteria.

The main result of MSPC is that apart from the four galaxies with powerful jets or radio lobes, their other source candidates have no special features that distinguish them from generic Seyferts and LINERs.  On that basis, they conclude that the UHECR-AGN correlation must be due to chance, {\em under the assumption that UHECR acceleration is not episodic in nature}.  

\section{Summary and Conclusions}

We have examined the 21 galaxies from the VCV catalog with $z \leq 0.018$ which are within 3.2$^\circ$ of the 27 highest-energy cosmic rays, and 8 additional AGNs from NED introduced by \citet{MSPC}.  We find that 14 of the 21 VCV galaxies are AGNs and the other seven are either star forming or quiescent galaxies.  X-ray or radio observations are needed to find out if any of these seven have obscured nuclear activity at an interesting level.  Two of the 8 additional NED galaxies can be optically established as AGNs (one is a Seyfert 2, one a LINER, and 6 do not have adequate optical spectra to make a determination).  The new optically identified AGN, WKK 4374, increases by one the number of UHECR-AGN correlations, because it is within 2.8$^\circ$ of a previously uncorrelated UHECR.   

We have determined the spectral type and bolometric luminosity of the 14 VCV AGNs and the newly identified AGN.  Their luminosities range from $5 \times 10^{42}$ erg s$^{-1}$ to 1 $\times$ $10^{46}$ erg s$^{-1}$. Five are broad line, nine are narrow-line and one is a BL Lac.  \citet{haoAGN05} finds that the numbers of broad line and narrow-line AGNs are comparable at low and moderate luminosity, while at high luminosity, broad line AGNs far outnumber narrow-line ones.  Thus the AGNs possibly producing the observed UHECRs are representative of the population of moderate luminosity AGNs both in luminosity and type.  

We have also identified and determined the bolometric luminosities of the Virgo AGNs and LINERs. Most have luminosities more than an order of magnitude below the lowest luminosity of any AGN correlated with a UHECR.  (A large number of low luminosity AGNs and LINERs are known in Virgo, due to the \citet{Ho1995} high-sensitivity spectroscopic survey of the near-by galactic nuclei in the northern hemisphere.)  A minimum bolometric luminosity requirement for UHECR-accelerating systems may therefore explain the absence of UHECRs correlated with Virgo AGNs.  If there is such a threshold, the significance of correlations between UHECRs observed by northern hemisphere observatories and VCV galaxies would be reduced in comparison to that observed by Auger South, due to a greater dilution of the VCV catalog in the northern hemisphere by large numbers of known very low luminosity AGNs.  A threshold in the bolometric luminosity is consistent with the Giant AGN Flare scenario \citep{FarrarandGruzinov}, but is not predicted by it, since the properties of the remnant have not yet been modeled.

Altogether, 21 of the 27 Auger UHECRs are within $3.2^\circ$ of a Veron-Cetty Veron galaxy or the newly identified AGN.  Of these, 17 of the 21 correlated UHECRs can be attributed to an identified AGN within $3.2^\circ$ and $z \leq 0.018$.  However few of the correlated AGNs satisfy the condition $L_{\rm bol} \geq 10^{45} E_{20}^2~{\rm erg \, s^{-1}}$, required to confine a cosmic ray as it is accelerated \citep{FarrarandGruzinov}.  If correlated AGNs with inadequate bolometric luminosities are in fact the sources of most of these UHECRs, their luminosities present a puzzle to the conventional picture of AGN acceleration in a continuous jet.  The recent analysis of \citet{MSPC} finds that few of the correlated AGNs have powerful radio jets or lobes, corroborating this conclusion.   If the AGN-UHECR correlation is a real one, these results provide evidence in favor of the new mechanism of UHECR acceleration by giant AGN flares proposed by \citet{FarrarandGruzinov}, in which a modest AGN has an intense flare producing a jet of the required luminosity (initiated for instance by a stellar tidal disruption rapidly heating the accretion disk) and then reverts to a mild-mannered existence.

\acknowledgements
We thank A. Berlind, A. Chou, J. Gelfand, J. Gunn, J. Huchra, I. Marquez-Perez, N. Masetti, J. Moustakas, D. Schneider, M. Strauss, and A. Tilak for valuable discussions and information.  GRF and IZ acknowledge the comments and encouragement of their colleagues in the Pierre Auger Collaboration.  This research has been supported in part by NSF-PHY-0701451.

\bibliographystyle{apj}
\bibliography{CR,AGN}

\appendix

\section{A. Notes on Individual Objects}
\label{section:individualobjects}
The 21 VCV galaxies correlated with Auger UHECRs with energy above 57 EeV are comprised of one BL Lac (Centaurus A), 5 broad line AGNs (Seyfert 1 galaxies), 8 narrow-line AGNs (Seyfert 2 galaxies) and 7 (non-optical-AGN) emission line galaxies. In this section we give detailed information on the identification and properties of each of these galaxies.  Work is underway to examine all of the VCV galaxies with $z \leq 0.024$, to determine their correct classifications and find the bolometric luminosities of those with active nuclei (IZ, GRF, N. Bursky-Tammam, E. Storm and A. Hoeft, in progress.)

%------------------------------Cen A------------------------------------------------------------------------------
\subsection{Centaurus A}
Centaurus A (NGC 5128) \citep{Israel1998} is a well known, nearby (3.4 Mpc) BL Lac. Its nuclear activity is manifested by large radio/X-ray jets - radio plumes extending to 250 kpc, and a compact circumnuclear disk. \citet{Rothschild2006} report a 2-10 keV X-ray flux which varies between 1.69 - 3.23 $\times$ 10$^{-10}$ erg s$^{-1}$ cm$^{-2}$ in six observations between August, 1996 and February, 2004 with the PCU2 instrument on {\it RXTE}.

%------------------------------Broad-Line AGN---------------------------------------------------------------------
\subsection{Broad-Line AGNs (Seyfert 1 Galaxies)}
{\bf SDSS J03302-0532 (J033013.26-053235.9)}: The optical spectrum shows broadened H$\alpha$ and H$\beta$ lines. We fit multi-Gaussian models to the spectrum. Galaxy subtraction is performed using the principle component analysis code of \citet{Hao2005}, and the narrow lines are modeled as outlined in \citet{GreeneandHo2004}; see also \citet{Ho1997c}. Following \citet{GreeneandHo2005}, the AGN luminosity is measured from the H$\alpha$ line and the velocity dispersion is measured from the H$\alpha$ line width (for details see \citet{GreeneandHo2007}). The black hole masses for the SDSS objects are estimated from the broad H$\alpha$ luminosity and line width using the relation $M_{\rm BH} = 2.0 \times 10^6 (\frac{L_{\rm H\alpha}}{10^{42} {\rm erg s^{-1}}})^{0.55} (\frac{{\rm FWHM_{H\alpha}}}{10^3 {\rm km s^{-1}}})^{2.06} M_{\odot}$ following \citet{GreeneandHo2005}. We find: FWHM(H$\alpha$) = 5340 km/s, log($L$(H$\alpha$)) = 40.83 erg s$^{-1}$, log($L$[OIII]) = 39.50 erg s$^{-1}$, $M_{\rm BH}$ = 7.2 $\times$ 10$^{7}$ $M_{\odot}$, $L_{bol}/L_{Edd}$ = 0.006. 

{\bf SDSS J03349-0548 (J033458.00-054853.2)}: The optical spectrum shows broadened H$\alpha$ and H$\beta$ lines. We fit the spectrum as described for SDSS J03302-0532 and find: FWHM(H$\alpha$) = 4390 km/s, log($L$(H$\alpha$)) = 40.10 erg s$^{-1}$, log($L$[OIII]) = 39.30 erg s$^{-1}$, $M_{\rm BH}$ = 6.6 $\times$ 10$^{7}$ $M_{\odot}$, $L_{bol}/L_{Edd}$ = 0.015. 

{\bf Zw 374.029 (CGCG 374-029)}: A broadened H$\alpha$ line can be seen in the optical spectrum with a broad H$\alpha$ FWHM of 2048 km/s \citep{Pietsch2000}. We have fit a single Gaussian to the narrow [OIII] line and find $F$[OIII] = 2.74 $\times$ 10$^{-14}$ erg s$^{-1}$ cm$^{-2}$. 

{\bf NGC 424}: The broad H$\alpha$ and H$\beta$ lines were detected in polarized emission by \citet{Moran2000}. They report an H$\alpha$ FWZI of approximately 12,000 km/s. However, since the broad lines are not detected in unpolarized emission, we use the narrow-line [OIII] flux, $F$[OIII] = 4267.0 $\times$ 10$^{-16}$ erg s$^{-1}$ cm$^{-2}$ \citep{Gu2006} to derive the bolometric luminosity.

{\bf 4U 1344-60}: \citet{Piconcelli2006} report a broad H$\alpha$ line with a FWHM of 4400 km/s. They classify it as an intermediate Seyfert (S1.5) due to the absence of a broadened H$\beta$ line. They report a de-reddened broad H$\alpha$ flux of $F$(H$\alpha$) = 3.3 $\pm$ 0.8 $\times$ 10$^{-12}$ erg s$^{-1}$ cm$^{-2}$, and a 2-10 keV X-ray flux of $F_{\rm 2-10keV}$ = 3.6 $\times$ 10$^{-11}$ erg s$^{-1}$ cm$^{-2}$.

%\clearpage
%----------------------------Narrow-Line AGN-------------------------------------------------------------------
\subsection{Narrow-Line AGNs (Seyfert 2 Galaxies)}

{\bf NGC 5506}: The optical spectrum for NGC 5506 has strong [OIII] and [NII] narrow lines. The log([OIII]/H$\beta$) and log([NII]/H$\alpha$) line ratios, 0.88 and -0.09 respectively \citep{IC5169}, place it firmly in the Seyfert 2 region of the BPT diagram. \citet{Gu2006} report an [OIII] flux, $F$[OIII] = 1614.0 $\times$ 10$^{-16}$ erg s$^{-1}$ cm$^{-2}$ and an X-ray luminosity, log($L_{\rm 2-10keV}$) = 42.89 erg s$^{-1}$. NGC 5506 also hosts circumnuclear H$_2$O masers.

{\bf IC 5135 (NGC 7130)}: The optical spectrum for IC 5135 has strong [OIII] and [NII] narrow lines. The log([OIII]/H$\beta$) and log([NII]/H$\alpha$) line ratios, 0.6990 and -0.0231 respectively \citep{Vaceli1997}, place it in the Seyfert 2 region of the BPT diagram. \citet{Shu2007} report an [OIII] luminosity, log($L$[OIII]) = 42.55 erg s$^{-1}$.

{\bf MRK 607 (NGC 1320)}: The log([OIII]/H$\beta$) and log([NII]/H$\alpha$) line ratios for MRK 607, 1.0458 and -0.1498 respectively \citep{Vaceli1997}, place it in the Seyfert 2 region of the BPT diagram. \citet{Shu2007} report an [OIII] luminosity, log($L$[OIII]) = 41.05 erg s$^{-1}$. MRK 607 also hosts circumnuclear H$_2$O masers.

{\bf IC 4518A}: We fit for the narrow lines in the optical spectrum of IC 4518A (courtesy of N. Masetti). The log([OIII]/H$\beta$) and log([NII]/H$\alpha$) line ratios (courtesy of J. Moustakas), 0.839 and -0.199 respectively, place it in the Seyfert 2 region of the BPT diagram. We fit a single Gaussian to the [OIII] line and obtain a flux of $F$[OIII] = 1.20 $\times$ 10$^{-13}$ erg s$^{-1}$ cm$^{-2}$.

{\bf NGC 1358}: The log([OIII]/H$\beta$) and log([NII]/H$\alpha$) line ratios, 1.0542 and 0.3032 respectively \citep{Ho1997a}, place it in the Seyfert 2 region of the BPT diagram. \citet{Shu2007} report an [OIII] luminosity, log($L$[OIII]) = 40.86 erg s$^{-1}$. The SMBH mass is determined to be 6.56 $\pm$ 2.50 $\times$ 10$^7$ $M_{\odot}$ \citep{Wu2001}. This implies an Eddington ratio of 0.029.

{\bf ESO 383-G18}: The log([OIII]/H$\beta$) and log([NII]/H$\alpha$) line ratios for ESO 383-G18, 0.9289 and -0.6327 respectively \citep{deGrijp1992}, place it in the Seyfert 2 region of the BPT diagram. \citet{Gu2006} report an [OIII] flux, $F$[OIII] = 360.9 $\times$ 10$^{-16}$ erg s$^{-1}$ cm$^{-2}$.

{\bf NGC 4945}: Optical emission from NGC 4945 is highly obscured. Even X-rays below 10 keV are absorbed by a column of N$_{\rm H}$ = 4.5 $\times$ 10$^{24}$ cm$^{-2}$, but NGC 4945 is one of the brightest Seyfert galaxies at 100 keV, indicating an active nucleus \citep{Madejski2000}. From 1-500 keV X-ray luminosity, \citet{Madejski2000} estimate the bolometric luminosity to be $\sim$2 $\times$ 10$^{43}$ erg s$^{-1}$. The mass of the SMBH has been determined from circumnuclear H$_2$O masers to be 
$\sim$1.4 $\times$ 10$^6$ $M_{\odot}$, which implies that the Eddington ratio for NGC 4945 is $\sim$0.1 \citep{Greenhill1997}.

{\bf ESO 139-G12}: The optical spectrum for ESO 139-G12 \citet{Marquez2004} does not extend to [OIII] and H$\beta$ wavelengths, but the hint of a broadened H$\alpha$ line indicates that it is an AGN. Since the spectrum is not flux calibrated, we use the upper limit for the X-ray luminosity from \citet{Polletta1996}, $L_{\rm 2-10keV}$ $<$ 47.56 $\times$ 10$^{42}$ erg s$^{-1}$.

%\clearpage
%-----------------------Non-AGN----------------------------------------------------------------------------------
\subsection{Non-AGN Emission Line Galaxies}

{\bf NGC 2989}: VCV identifies NGC 2989 as an HII galaxy. \citet{NGC2989} (NB: Co-authors are Veron-Cetty and Veron) classify this as a pure HII galaxy, based on flux ratios, log([OIII]/H$\beta$) =  -0.2304, log([NII]/H$\alpha$) = -0.2757. This falls within but almost on the theoretical star formation line developed by \citet{IC5169}. It may be a composite AGN-Starburst (transitioning) object with AGN contribution to the total luminosity $\leq$ 20\%.

{\bf NGC 7591}: The log([OIII]/H$\beta$) and log([NII]/H$\alpha$) line ratios, -0.3455 and -0.3224 respectively \citep{Moustakas2006}, place it below the star formation line determined by \citet{Kauffmann2003}. The infrared luminosity, Log($L_{IR}$) = 11.12 $L_{\odot}$, indicates that it is a luminous infrared galaxy (LIRG). \citet{Sturm} classifies it to be a Starburst galaxy while \citet{Veilleux} classifies it as a LINER, resulting from shocks and not from AGN accretion activity.

{\bf NGC 1204}: The (de-reddened) log([OIII]/H$\beta$) and log([NII]/H$\alpha$) line ratios for NGC 1204, -0.47 and -0.11 respectively \citep{Veilleux}, place it between the \citet{Kauffmann2003} and \citet{IC5169} lines in the LINER region. \citet{Veilleux} give reasons to believe that the emission lines are a result of shocks instead of accretion. \citet{Sturm} also classify it as a star forming galaxy. However, \citet{Corbett2003}, based on their radio and spectroscopic COLA survey, claim this galaxy lies on the border between Seyfert and LINER and classify it as a Seyfert even though no radio core was detected and the radio continuum is dominated by star formation. They place a limit on the AGN component, of 25\% of the total emission of the galaxy.

{\bf NGC 5244}: While the optical spectrum for NGC 5244 \citep{NGC5244} has [NII] and H$\alpha$ emission lines, [OIII] and H$\beta$ lines are not apparent. \citet{NGC5244} classifies it as an HII galaxy and VCV also identifies it as an HII galaxy.

{\bf IC 5169}: \citep{IC5169} classifies this as an HII galaxy based on the following line ratios, log([OIII]/H$\beta$) = 0.20, log([NII]/H$\alpha$) = -0.170, log([OI]6300/H$\alpha$) = -1.49, log([SII]6717,31/H$\alpha$) = -0.55. While it falls between the \citet{Kauffmann2003} and \citet{IC5169} lines in the log([OIII]/H$\beta$) vs. log([NII]/H$\alpha$) diagram, it falls well within the star formation line in log([OIII]/H$\beta$) vs. log([SII]/H$\alpha$) and log([OIII]/H$\beta$) vs. log([OI]/H$\alpha$) diagrams, confirming that it is not an optical AGN.

{\bf NGC 7135}: \citet{NGC7135} classify NGC 7135 as a star forming galaxy based on the fact that the [OIII] emission line is very weak, EW([OIII]) = -0.7A, and that strong stellar absorption features are visible, determining it to be ``an early-type galaxy with no active nucleus''.

{\bf Q 2207+0122 (PC 2207+0122)}: Although the optical spectrum for Q 2207+0122 (courtesy of D. Schneider) shows emission lines, Q 2207+0122 is just an emission line galaxy and not an AGN (confirmed by D. Schneider, private communications). The log([OIII]/H$\beta$) and log([NII]/H$\alpha$) line ratios (courtesy of J. Moustakas), 0.815 and -1.581 respectively, place in the star formation region \citet{Kauffmann2003}. This galaxy is a low metallicity dwarf galaxy (J. Moustakas, private communication).  
%\clearpage

%-----------------------------------------------------------------------------------------------------------------------
\section{B. Details on the AGNs in the Virgo Cluster}
\label{section:VirgoDetails}
The details for the VCV galaxies in the Virgo Cluster are listed in the table below, in order of distance from the center of the cluster at $\alpha_{\rm J2000}$ = 12:26:32.1, $\delta_{\rm J2000}$ = +12:43:24 \citep{Virgo}. The [OIII] and 2-10 keV X-ray luminosities, as well as the bolometric luminosities derived from them, are given for the AGNs not listed as HII galaxies in VCV. Note that the bolometric luminosities derived from [OIII] luminosities are systematically higher than those derived from 2-10 keV X-ray luminosities. There is some evidence that the [OIII] to bolometric luminosity conversion depends on luminosity or Eddington ratio (e.g., \citet{Netzer2006}).  Since optical emission line fluxes, particularly H$\alpha$, may be contaminated by non-nuclear sources (e.g., \citet{Ho2001}) we have chosen to use the hard X-ray flux as a bolometric indicator whenever possible. 
\begin{table}[h!] 
\caption{Details on the Virgo AGNs in VCV \label{VirgoLum}}
\begin{tabular}{cccccccccc}
\hline
\hline
Name & RA & Dec & z & VCV Class & Ho1997Class & Log($L$[OIII]) & Log($L_{\rm 2-10 keV}$) & $L_{bol,{\rm [OIII]}}$ & $L_{bol,{\rm 2-10keV}}$ \\
    & (J2000) & (J2000) & & & & erg s$^{-1}$ & erg s$^{-1}$ & erg s$^{-1}$ & erg s$^{-1}$ \\
\hline
NGC 4388 & 186.44 & $+$12.66 & 0.008 & S1h & S1.9 & 40.35 & 41.72$^{a}$ & 7.85 $\times$ 10$^{43}$ & 1.05 $\times$ 10$^{43}$ \\
NGC 4374 & 186.27 & $+$12.89 & 0.003 & S2  & L2 &   38.45 & 39.12$^{b}$ & 9.78 $\times$ 10$^{41}$ & 2.64 $\times$ 10$^{40}$ \\
NGC 4438 & 186.94 & $+$13.01 & 0.004 & S3b & L1.9 & 38.80 & 39.22$^{c}$ & 2.22 $\times$ 10$^{42}$ & 3.32 $\times$ 10$^{40}$ \\
NGC 4486 & 187.71 & $+$12.39 & 0.004 & S3  & L2 &   39.07 & 40.75$^{c}$ & 4.15 $\times$ 10$^{42}$ & 1.12 $\times$ 10$^{42}$ \\
NGC 4477 & 187.51 & $+$13.64 & 0.005 & S2  & S2 &   38.82 & 39.65$^{a}$ & 2.30 $\times$ 10$^{42}$ & 8.93 $\times$ 10$^{40}$ \\
NGC 4501 & 188.00 & $+$14.42 & 0.007 & S2  & S2 &   39.10 & 39.59$^{a}$ & 4.38 $\times$ 10$^{42}$ & 7.78 $\times$ 10$^{40}$ \\
NGC 4552 & 188.92 & $+$12.56 & 0.001 & S2  & T2 &   38.05 & 39.55$^{d}$ & 3.94 $\times$ 10$^{41}$ & 7.10 $\times$ 10$^{40}$ \\
NGC 4550 & 188.88 & $+$12.22 & 0.001 & S3  & L2 &   38.15 & $<$ 37.68$^{d}$ & 4.95 $\times$ 10$^{41}$ & $<$ 9.57 $\times$ 10$^{38}$ \\
NGC 4569 & 189.21 & $+$13.16 & 0.004 & S   & T2 &   39.27 & 39.41$^{b}$ & 6.54 $\times$ 10$^{42}$ & 5.2 $\times$ 10$^{40}$ \\
NGC 4548 & 188.86 & $+$14.51 & 0.002 & S3  & L2 &   38.11$^{f}$ & --- & 4.54 $\times$ 10$^{41}$ & --- \\
NGC 4579 & 189.43 & $+$11.82 & 0.005 & S3b & S1.9/L1.9 & 39.42 & 41.15$^{c}$ & 9.25 $\times$ 10$^{42}$ & 2.83 $\times$ 10$^{42}$ \\
NGC 4168 & 183.07 & $+$13.21 & 0.008 & S1.9 & S1.9 &37.91 & 39.87$^{a}$ & 2.84 $\times$ 10$^{41}$ & 1.48 $\times$ 10$^{41}$ \\
NGC 4383 & 186.36 & $+$16.47 & 0.005 & H2 & --- & --- & --- & --- & --- \\
NGC 4192 & 183.45 & $+$14.90 & 0.004 & S3 & T2 &    38.12 & 39.58$^{e}$ & 4.57 $\times$ 10$^{41}$ & 7.6 $\times$ 10$^{40}$ \\
NGC 4639 & 190.72 & $+$13.26 & 0.001 & S1.0 & S1.0 &38.41 & 40.22$^{a}$ & 8.96 $\times$ 10$^{41}$ & 3.32 $\times$ 10$^{41}$ \\
NGC 4450 & 187.12 & $+$17.08 & 0.006 & S3b & L1.9 & 38.78 & 40.35$^{e}$ & 2.10 $\times$ 10$^{42}$ & 4.46 $\times$ 10$^{41}$ \\
NGC 4472 & 187.45 &  $+$8.00 & 0.003 & S2 & S2 &    37.81 & $<$ 39.32$^{a}$ & 2.25 $\times$ 10$^{41}$ & $<$ 4.18 $\times$ 10$^{40}$ \\
NGC 4694 & 192.06 & $+$10.98 & 0.004 & H2 & HII & --- & --- & --- & --- \\
IC 3576  & 189.16 &  $+$6.62 & 0.003 & H2 & --- & --- & --- & --- & --- \\
NGC 4698 & 192.10 &  $+$8.49 & 0.003 & S2 & S2  &   38.81$^{f}$ & --- & 2.25 $\times$ 10$^{42}$ & --- \\
NGC 4303 & 185.48 &  $+$4.47 & 0.005 & S2 & HII/L$^{g}$ &   39.13  &  39.16$^{g}$  &  4.72 $\times$ 10$^{42}$ &  2.89 $\times$ 10$^{40}$ \\
NGC 4412 & 186.65 &  $+$3.97 & 0.008 & S2 &  S2$^{h}$  & --- & --- & --- & ---- \\              
NGC 3976 & 178.99 &  $+$6.75 & 0.008 & S2 & S2  &   39.12  &   ---   &  4.61 $\times$ 10$^{42}$ &    ---  \\
NGC 4636 & 190.71 &  $+$2.69 & 0.003 & S3b & L1.9 & 38.10  &  39.25$^{c}$  &  4.41 $\times$ 10$^{41}$ & 3.56 $\times$ 10$^{40}$ \\  
MRK 52   & 186.43 &  $+$0.57 & 0.007 & H2  & --- &   ---   &   ---   &  ---   &   --- \\   
NGC 4772 & 193.37 &  $+$2.17 & 0.003 & S3b & L1.9 &  38.45 &   ---   &  9.86 $\times$ 10$^{41}$ &    ---  \\
SDSS J12122+0004  & 183.06 & $+$0.07 &  0.008 & S1 & HII & --- & --- & --- & --- \\
(MRK 1313) & & & & & & & & & \\
NGC 4418 & 186.73 &  $-$0.88 & 0.007 & S2 &  S2$^{h}$  &  ---  & 39.25$^{i}$ &  --- & 3.56 $\times$ 10$^{40}$  \\
SDSS J12580+0134  & 194.51  & $+$1.58 &  0.004 &  S2 &  HII & 38.19 & --- & 5.45 $\times$ 10$^{41}$  & ---  \\
(NGC 4845) & & & & & & & & & \\
MRK 1308 & 178.55 &  $+$0.14 & 0.004 & H2 & --- &  ---  &  ---  & --- &  --- \\
\hline
\hline
\end{tabular}
\tablecomments{S=Seyfert, L=Liner, T=Transition object. The classifications in column 6 and the [OIII] luminosities in column 7 (except where marked) are taken from \citet{Ho1997a}.} 
\tablerefs{(a) \citet{Panessa2006}, (b) \citet{Ho2001}, (c) \citet{Gonzalez2006}, (d) \citet{Satyapal2005}, (e) \citet{Terashima2002}, (f) \citet{Shields2007}. (g) \citet{Horst2008}, (h) \citet{IC5169}, (i) \citet{Kandalyan2005}}
\end{table}

\section{C. Notes on the AGNs from the NASA Extragalactic Database}
\label{section:MoskalenkoAGNs}
\citet{MSPC} lists eight additional AGNs within 3.2$^{\circ}$ of the Auger CRs with $z \leq 0.018$, based on the classifications in the NASA Extragalactic Database. Some of them were determined to be AGNs after the publications of the VCV 12th Edition. Others were selected as AGNs mainly via X-ray, infrared, or radio criteria and show only some evidence of nuclear activity in the optical, which may explain their absence from VCV. One of these AGNs, WKK 4374, correlates with a CR (Auger year = 2007, Auger day = 186; E=64 EeV),  which was previously uncorrelated. Furthermore, two of the additional AGNs, NGC 7626 and NGC 2907, correlate with two CRs, (2006, 185; E=83 EeV) and (2007, 84; E=64 EeV), respectively, whose VCV counterparts were shown in this paper to be non-nuclear dominated emission line galaxies. We perform a literature search on the extra AGNs and list their optical properties below. 

{\bf WKK 4374 (IGR J14515-5542)}: From optical spectra of INTEGRAL sources, \citet{Masetti2006} determined WKK 4374 to be a Seyfert 2 galaxy. They report an absorption-corrected [OIII] flux of $6.4 \pm 1.6 \times 10^{-14}$ erg s$^{-1}$ cm$^{-2}$. This gives an [OIII] luminosity of $5.4 \times 10^{40}$ erg s$^{-1}$ and implies a bolometric luminosity of $1.9 \times 10^{44}$, within the range found for the correlated AGNs from VCV; $\lambda_{\rm bol} = 0.46$.

{\bf NGC 5140 and NGC 2907}: These two galaxies were determined to have narrow LINER-like emission by \citet{Mauch2007} from visual inspections of the 6dFGS optical spectra. No emission line fluxes or ratios have been published.

{\bf NGC 7626}: This is a radio loud galaxy. \citet{Ho1997a} determine it to have a possible L2 nucleus (the classification is characterized as highly uncertain), with an [OIII] luminosity of $3.3 \times 10^{38}$ erg s$^{-1}$ (100\% uncertainty), which implies a bolometric luminosity of $1.2 \times 10^{42}$ erg s$^{-1}$, giving $\lambda_{\rm bol} = 0.0017$.  Since some correlated galaxies are chance coincidences, this may be one of them.

{\bf PKS 1343-60 (Centaurus B)}: Cen B is a well-known FRI radio galaxy. However, while an optical counter-part for the active nucleus has been detected \citep{West1989}, the optical emission is highly absorbed and no clear emission lines or line ratios could be obtained from the optical spectrum beyond the detection of the H$\alpha$-[NII] complex.

{\bf NGC 5064}: \citet{Bonatto1989} report a [NII]/H$\alpha$ value of 1.051 and [SII]/H$\alpha$ value of 0.461. No [OII], H$\beta$, or [OI] line fluxes were reported. This indicates that NGC 5064 is a LINER. The authors determine this LINER to be nuclear powered rather than shock powered.

{\bf IRAS 13028-4909}: This galaxy is from a set of X-ray and IR selected AGNs \citep{Kirhakos1990}. An [NII]/H$\alpha$ ratio of 1.28 was reported. No other line fluxes or ratios were given. 

{\bf AM 1754-634 NED03}: Shows a bright optical nucleus. However \citet{Fairall1981} reports that the 5007 and 4954 [OIII] lines were of equal strength (instead of in a 3:1 ratio), H$\alpha$ was barely detected, and no other emission lines were detected. This casts doubt on the strength of its nuclear activity based on optical data. Data from other wavelengths has not been published.

\end{document}